\documentclass[12pt]{article}

\usepackage{amssymb,epsfig}
\usepackage{multirow} 

\begin{document}

\begin{center}
\baselineskip=24pt

{\Large \bf Simulations of neutron background in a time projection
  chamber relevant to dark matter searches}
\vspace{0.5cm}

\baselineskip=24pt

{\large
M. J. Carson, J. C. Davies\footnote{Corresponding author, 
e-mail: j.c.davies@sheffield.ac.uk}, E. Daw, R. J. Hollingworth,\\
J. A. Kirkpatrick, V. A. Kudryavtsev, T. B. Lawson, P. K. Lightfoot,\\
J. E. McMillan, B. Morgan, S. M. Paling, M. Robinson,\\
N. J. C. Spooner, D. R. Tovey, E. Tziaferi}
\vspace{0.5cm}

{\it Department of Physics and Astronomy, 
University of Sheffield, Sheffield S3 7RH, UK}

\vspace{0.5cm}
\begin{abstract}
Presented here are results of simulations of neutron
background performed for a time projection chamber acting as a
particle dark matter detector in an underground laboratory. The
investigated background includes neutrons from rock and detector
components, generated via spontaneous fission and ($\alpha$,n)
reactions, as well as those due to cosmic-ray muons. 
Neutrons were propagated to the sensitive volume of the detector
and the nuclear recoil spectra were calculated.
Methods of neutron background suppression were also examined and 
limitations to the sensitivity
of a gaseous dark matter detector are discussed. Results indicate that
neutrons should not limit sensitivity to WIMP-nucleon interactions 
down to a level of $(1-3) \times 10^{-8}$ pb in a 10 kg detector.

\end{abstract}

\end{center}

\vspace{0.5cm}


\vspace{0.5cm}
\noindent Corresponding author: J. C. Davies,
Department of Physics and Astronomy, University of Sheffield, 
Hicks Building, Hounsfield Road, 
Sheffield S3 7RH, UK

\noindent Tel: +44 (0)114 2223547; \hspace{2cm} Fax: +44 (0)114 2728079; 

\noindent E-mail: j.c.davies@sheffield.ac.uk

\pagebreak

{\large \bf 1. Introduction}
\vspace{0.3cm}

It is important to the sensitivity of future underground dark matter
detectors to have accurate simulations of background events and to find ways to
suppress or reject them. Presented here are simulations of a gas Time
Projection Chamber (TPC) as a WIMP (Weakly Interacting Massive
Particle) dark matter detector and the neutron background that may be
expected from the detector itself and the surrounding environment. A
time projection chamber is well suited for dark matter detection
because it enables directional track information to be recorded, and
so can be used to distinguish between postulated halo models
\cite{halo}. Low pressure is necessary for a gas detector to maximise
its spatial resolution. An example of this type of
detector is DRIFT I \cite{smpdrift}.

Neutron background is an important issue, since neutrons can produce
nuclear recoils in the same way as WIMPs. Other backgrounds
that have the capability to generate false signals include sources of
alphas and gammas. Simulations of these backgrounds, however, have
been neglected in this study as any signals generated by alpha
particles in the detector can be actively vetoed because they will be
produced at the edge of the detector and have long tracks, and those
due to gammas can be rejected by adjusting the energy threshold and
applying appropriate cuts \cite{smpdrift,dan,jakdrift}.

The simulations described use either the GEANT4 toolkit \cite{g4} or
FLUKA \cite{fluka} (versions FLUKA-2002 and FLUKA-2003) 
to simulate geometry, particle production,
tracking and detection, with the neutron production energy spectra 
from spontaneous fission and ($\alpha$, n) reactions being
calculated using SOURCES \cite{sources}. The background
investigated is that of neutrons produced via spontaneous fission and
($\alpha$, n) reactions in and around the detector, along with
muon-induced neutrons from deeply penetrating cosmic rays. The
neutrons are then allowed to undergo elastic and inelastic scattering which, if
within the fiducial volume of the detector, can produce nuclear
recoils that are indistinguishable from the WIMP-induced nuclear
recoils being searched for.  These simulations, therefore, attempt to
determine the rate of nuclear recoils that will be observed by the
detector due to the neutron background, to estimate the amount of
passive neutron shielding required to keep this rate sufficiently
suppressed and to formulate requirements for the purity of materials
used in the detector construction in order to gain a high sensitivity
to WIMPs.

Other background simulations for underground dark matter detectors
have been performed or are underway, including, for example, those for
a large-scale xenon detector \cite{nxenon}, CRESST-II at the Gran
Sasso Laboratory \cite{cresst}, the EDELWEISS-II experiment in the
Modane Underground Laboratory \cite{edelweiss}, the IGEX germanium
dark matter experiment at Canfranc \cite{igex} and the CsI detectors
at the CPL underground site in South Korea \cite{cpl}. 

In contrast to the aforementioned studies, however, the simulations
described here are for a low-pressure gas TPC detector. The
probability for a neutron or gamma interaction in such a detector is
very low and the light charged particles (i.e. electrons
and muons) are hardly visible. No multiple recoil events from neutrons
are expected in gas, and muons and muon-induced secondary particles
may not be seen, unlike the case for other experiments. This makes the
simulations for low-pressure gas detectors unique. The results of
other simulations cannot be extrapolated to such TPCs.
Some studies of neutron background in a gas TPC using different
simulation strategy and code have recently been completed by
Smith et al. \cite{pfssim}.

\vspace{0.5cm}
{\large \bf 2. Geometry}
\vspace{0.3cm}

\begin{figure}
\begin{center}
\epsfig{figure=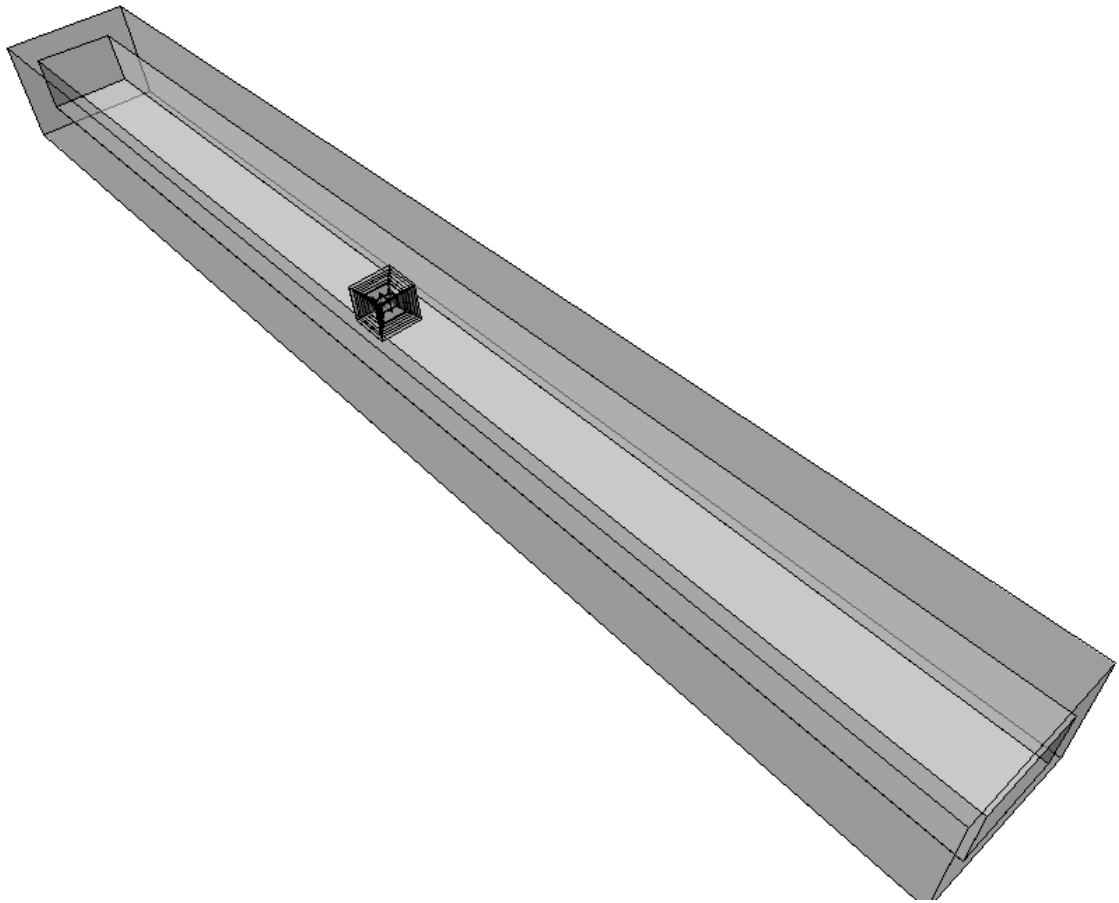,width=15cm}
\caption{GEANT4 image, using VRML, of the whole geometry used in the
  simulations.  The laboratory cavern can be seen surrounded by a
  volume of rocksalt.  The detector is positioned centrally within the
  laboratory hall.}\label{singleModule}
\end{center}
\end{figure}

For the GEANT4 \cite{g4} simulations a laboratory area  $80$ m long,
similar in size to the Boulby Underground Laboratory 
(North Yorkshire, UK; Boulby mine is run by Cleveland Potash
Ltd.), filled with air approximated at $1$ atmosphere of an $80:20$
nitrogen to oxygen mixture, was used (see Figure \ref{singleModule} for
illustration). Surrounding the laboratory is rock salt (pure NaCl in these
simulations) $3$ m thick on $4$ sides. The `tunnel ends' were left
open however, since the rock far from the detector,
including the end walls, contributes very little to the neutron flux
at the detector. Further reduction of the length of the simulated
laboratory in the simulations to $40$ m results in a less than $10\%$
decrease in the number of neutrons observed in the detector placed in
the middle of the cavern. Thus, to save CPU time, the cavern  length
was limited to $80$ m.

\begin{figure}
\begin{center}
\epsfig{figure=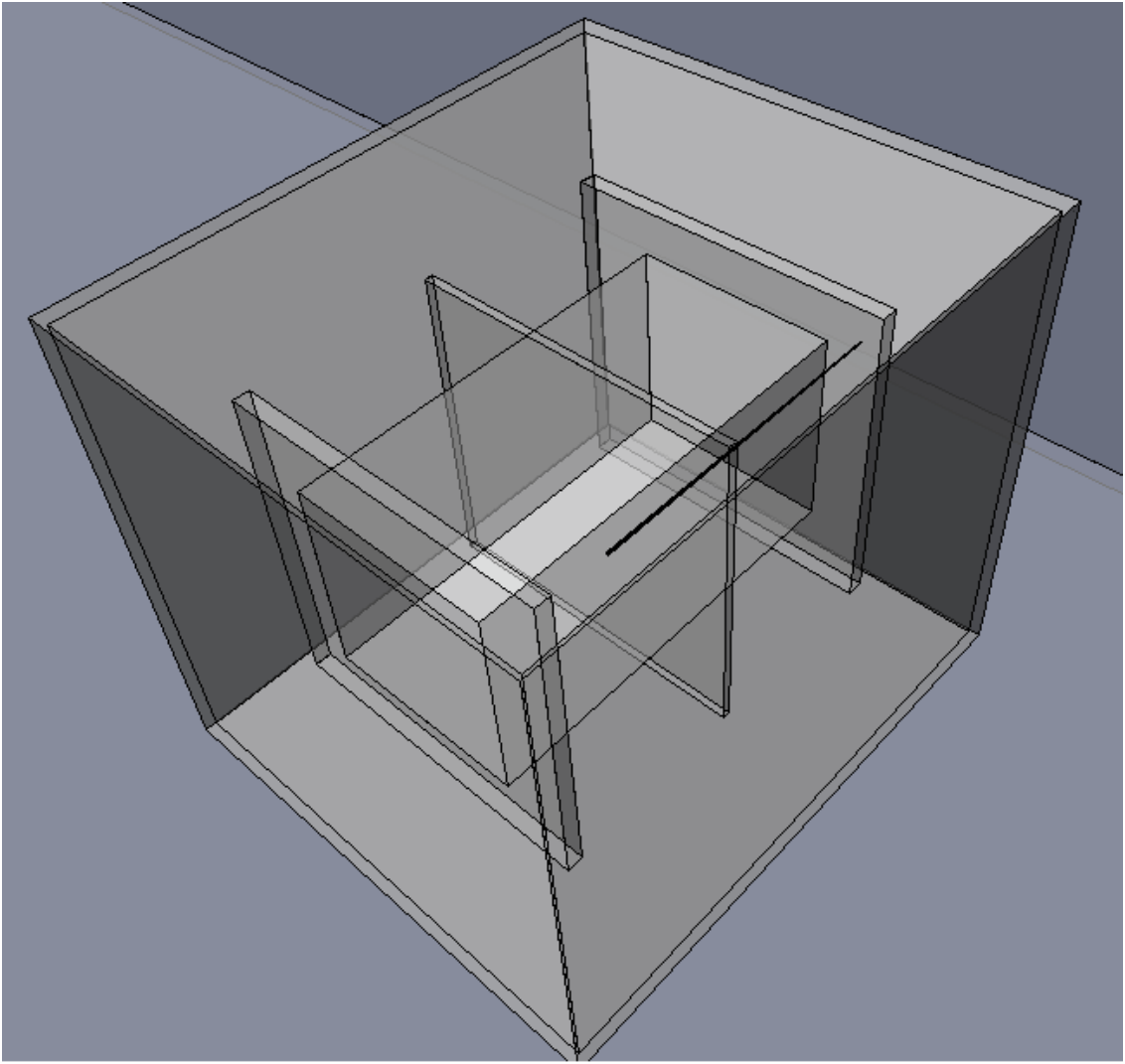,width=15cm}
\caption{GEANT4 image, using VRML, of the detector geometry used for the
  simulations.  The steel vessel can be seen surrounding the gas
  volume.  Inside, the perspex frames of the MWPCs and the cathode are
  positioned with the fiducial volume within, and the resistor volume
  can be seen along the side of them.}\label{detector}
\end{center}
\end{figure}

The detector itself consists of a $2$ cm thick stainless steel vessel
with internal dimensions of $1.2 \times 1.2 \times 1.4$ m$^{3}$ and a
mass of $1590$ kg (Figure \ref{detector}). The $2$ cm thickness was
used as an average, although a real detector of this type, such as the
DRIFT I detector \cite{smpdrift}, has thinner vessel walls with
support struts and thicker flanges to the front and rear of the
vessel. Of more importance in this approximation is the realistic
total mass of the vessel since this effects the neutron
production rate in the steel. Within the vessel the detector includes
a cathode frame with dimensions $0.8 \times 0.8 \times 0.02$ m$^{3}$
made of perspex and mounted vertically in the centre of the detector
volume, as suggested  (as one of the possible designs) for a future
larger scale TPC \cite{smpla}.  To either side of this frame are
another two frames, which are used to support the two multi-wire
proportional chambers (MWPCs).  The dimensions of these frames are
$0.8 \times 0.8 \times 0.04$ m$^{3}$ and they are positioned so the
inner grid wire plane sits $0.5$ m out to the edge from the central
cathode plane. The wires themselves have not been included, since the
increased complexity of the detector geometry would increase the
simulation run-time significantly yet give no improvement to the
overall results, as the mass of these wires and their effect are
negligible when compared to other detector components. The rest of the
volume within the vessel is filled with CS$_{2}$ gas at a pressure of
$160$ torr, with a density of $0.668$ kg m$^{-3}$. CS$_{2}$ gas is
chosen for DRIFT-type detectors as it is electronegative,
i.e. CS$_{2}$ molecules attract free electrons and drift in an
electric field towards the MWPC reducing diffusion, which is quite
large for drifting electrons \cite{cjmdrift}. The gas pressure ($160$
torr) was chosen to reduce the amount of diffusion of the ions in the
gas and therefore ensure good spatial resolution and
directionality. The fiducial volume within the MWPC frames is $0.5
\times 0.5 \times 1$ m$^{3}$, which gives a target mass of $0.167$ kg.
This is a possible design for the second generation DRIFT-type
detectors \cite{smpla}.

An approximation of the resistor chain in the field cage of the
detector is also included to simulate the neutron background from the
ceramic material of the resistors. The dimensions of the resistors were
approximated to a cuboid of ceramic material positioned along the long
side of the fiducial volume just outside the perspex MWPC frames.  The
volume of ceramics was $3.28 \times 3.28 \times 700 \rm\, mm^{3}$.

Varying amounts of passive neutron shielding (CH$_{2}$) were added to
the simulation geometry in order to evaluate the quantity required to
reduce the rate of nuclear recoils due to the rock neutron background
to less than one per year for a target mass of $3.33$ kg ($20$ of the
modules described above). This is a reasonable reduction that will
allow a sensitivity comparable to other world-class dark matter
detectors to be achieved.

\begin{figure}
\begin{center}
\epsfig{figure=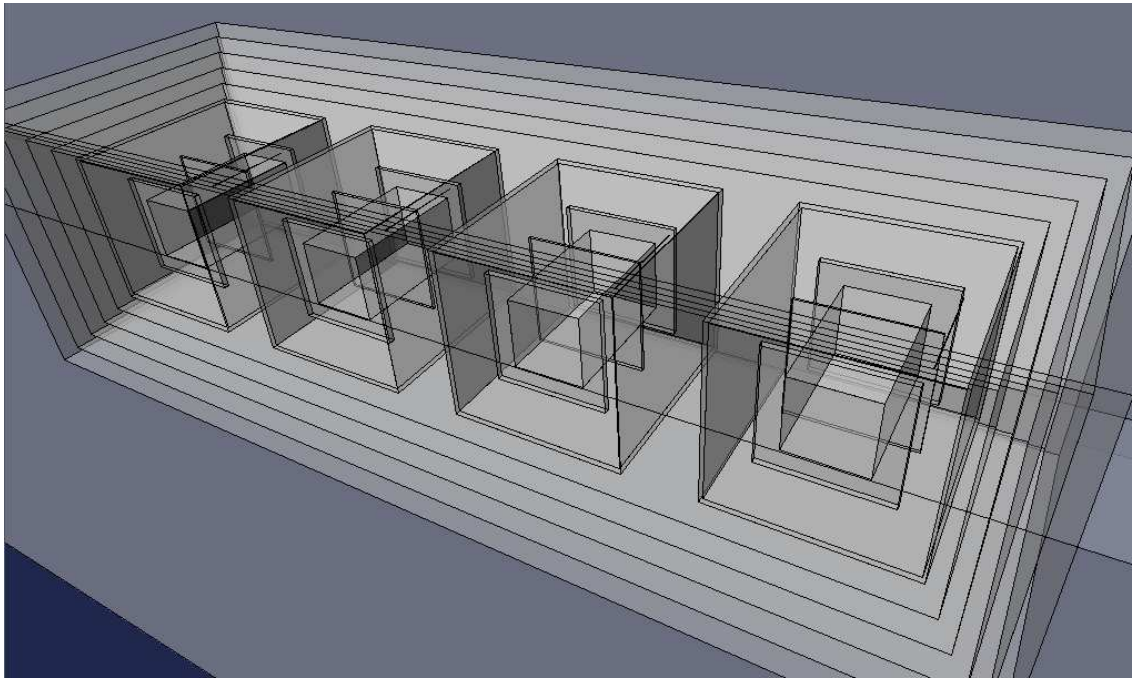,width=15cm}
\caption{GEANT4 image, using VRML, showing the geometry for
simulating 4 detectors.  The detectors are positioned centrally
within the laboratory hall and have CH$_{2}$ neutron shielding
both in between the vessels and surrounding the group of
vessels.}\label{4modulesFig}
\end{center}
\end{figure}

Multiple detectors were also simulated to assess the effects of
other neutron background sources and shielding between detectors.  To
investigate this, four detectors were grouped and positioned in a row
with varying amounts of CH$_{2}$ neutron shielding between the vessels
(see Figure \ref{4modulesFig} for illustration).

\vspace{0.5cm}
{\large \bf 3. Neutrons from rock}
\vspace{0.3cm}

Simulations of neutrons being produced in the rock surrounding the
detector laboratory were performed chiefly in order to establish the
amount of hydrocarbon shielding required to reduce the recoil rate
observed in the detector to well below one per year. This is
fundamental to gaining greater sensitivity and enhancing capabilities
of observing WIMP-nucleus interactions. 

Neutron production by radioactive isotopes in the decay chains of
uranium and thorium was calculated using the SOURCES code
\cite{sources}, as follows. Spontaneous fission (of $^{238}$U mainly)
was simulated using the Watt spectrum \cite{watt}. Neutron fluxes and
spectra from ($\alpha$,n) reactions were obtained taking into account
the lifetimes of isotopes, energy spectra of alphas, cross-sections of
reactions as functions of alpha energy, branching ratios for
transitions to different excited states, stopping power of alphas in
various media, and assuming isotropic emission of neutrons in the
centre-of-mass system.

The original SOURCES code provides a treatment of ($\alpha$,n)
reactions up to $6.5$ MeV $\alpha$-energies only. The code was
modified to allow treatment of alphas with higher energies
\cite{nxenon}. Existing cross-sections were extended to $10$ MeV,
taking into account available experimental data. For some materials
used in detector construction, new cross-sections were added to the
code. The Nuclear Data Services of the Nuclear Data Centre at the
International Atomic Energy Agency \cite{iaea} were used to obtain
cross-sections. The branching ratios for transitions to the ground and
excited states above $6.5$ MeV were chosen to be the same as at $6.5$
MeV. This resulted in a  small overestimate of neutron energies for
alphas above $6.5$ MeV, since the increased probability of transition
to the higher states was neglected.  If the excited levels were not in
the code library, as was the case for elements for which the
cross-sections were absent too, then in adding the cross-section we
assumed that the transition was occurring to the ground state only.

\begin{figure}
\begin{center}
\epsfig{figure=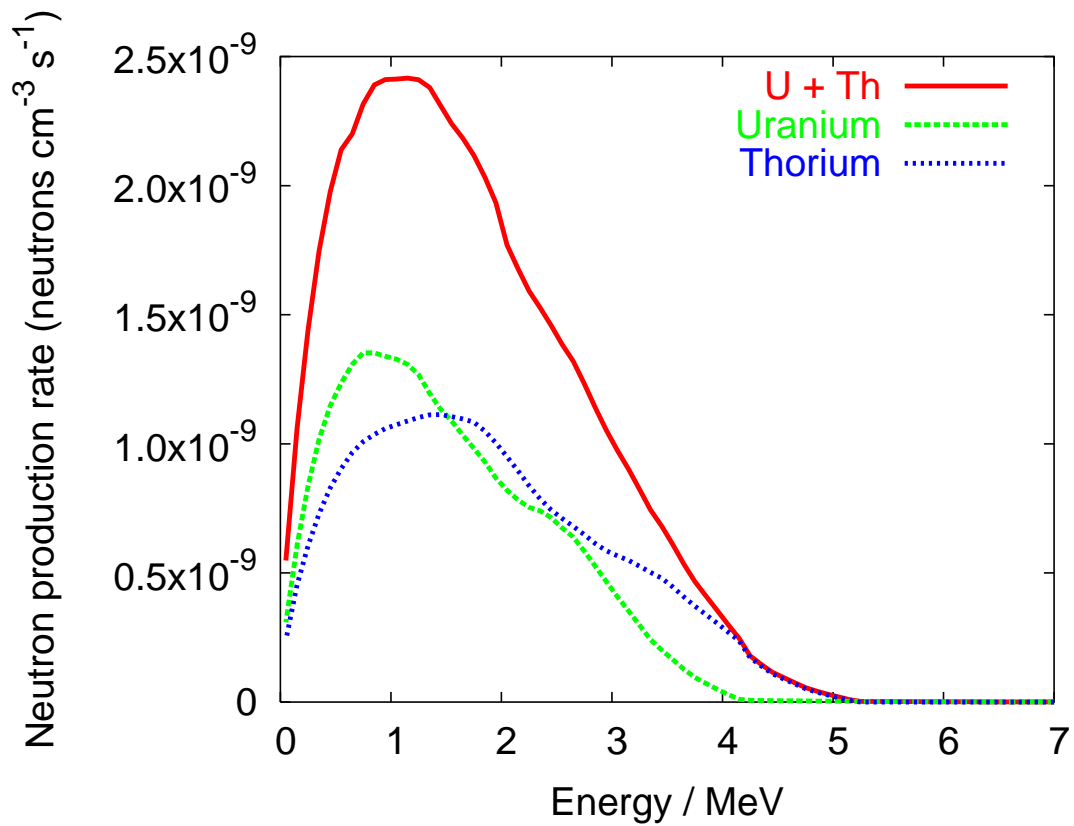,height=15cm,angle=-90}
\caption{Simulated energy spectrum of neutrons produced in the rock salt
  surrounding the detector, as calculated using modified SOURCES.
  Spectra due to $60$ ppb uranium contamination, $130$ ppb thorium
  contamination and U and Th combined are shown.  Solid curve -
  combined total neutron production rate due to uranium and thorium
  contamination in rock salt; dashed line - neutron production rate
  due to uranium; dotted line - neutron production rate due to
  thorium.}\label{InputSpecFig}
\end{center}
\end{figure}

The neutron energy spectrum produced by this modified SOURCES code
(shown in Figure \ref{InputSpecFig}) was assigned to neutrons being
generated isotropically within a $2.5$ m thickness of rock at the edge
of the laboratory volume. Values for the uranium and thorium content
in the rock ($60$ ppb and $130$ ppb, respectively) were taken from
measurements \cite{pfssim,ukdmc} at Boulby mine. This gives a total
neutron production rate of $6.32 \times 10^{-8}$ neutrons cm$^{-3}$
s$^{-1}$ with a mean energy of $1.73$ MeV. From this input spectrum
the neutrons were propagated through the rock to the laboratory cavern
using the GEANT4 toolkit. At the rock-laboratory boundary the neutron
positions and 4-momenta were recorded and the neutrons were
terminated. The flux of neutrons at the rock-lab boundary was found to
be $2.45 \times 10^{-6}$ cm$^{-2}$ s$^{-1}$ above $10$ keV, $1.94
\times 10^{-6}$ cm$^{-2}$ s$^{-1}$ above $100$ keV and $9.15 \times
10^{-7}$ cm$^{-2}$ s$^{-1}$ above $1$ MeV. These values are lower than
reported in Ref. \cite{nxenon} because of the smaller contamination levels
of U and Th used in the work presented here. The spectrum at this boundary
was then used to begin the next stage of neutron propagation, from the
rock-laboratory boundary to the detector. The volume of rock
surrounding the laboratory volume was kept during this second stage to
ensure that back-scattering of neutrons from the cavern walls was
allowed to occur. The detector was simulated both without any
shielding and also with varying thicknesses of hydrocarbon neutron
shielding. In the case of the shielded detector, the neutrons were
propagated through the shielding in several stages ($5$ g cm$^{-2}$
each stage) -- a convenient technique to reduce the CPU time, improve
statistics and, therefore, decrease statistical uncertainties. In all
runs the number of events processed was large enough to ensure the
statistical uncertainties were $\ll 1 \%$.

\begin{figure}
\begin{center}
\epsfig{figure=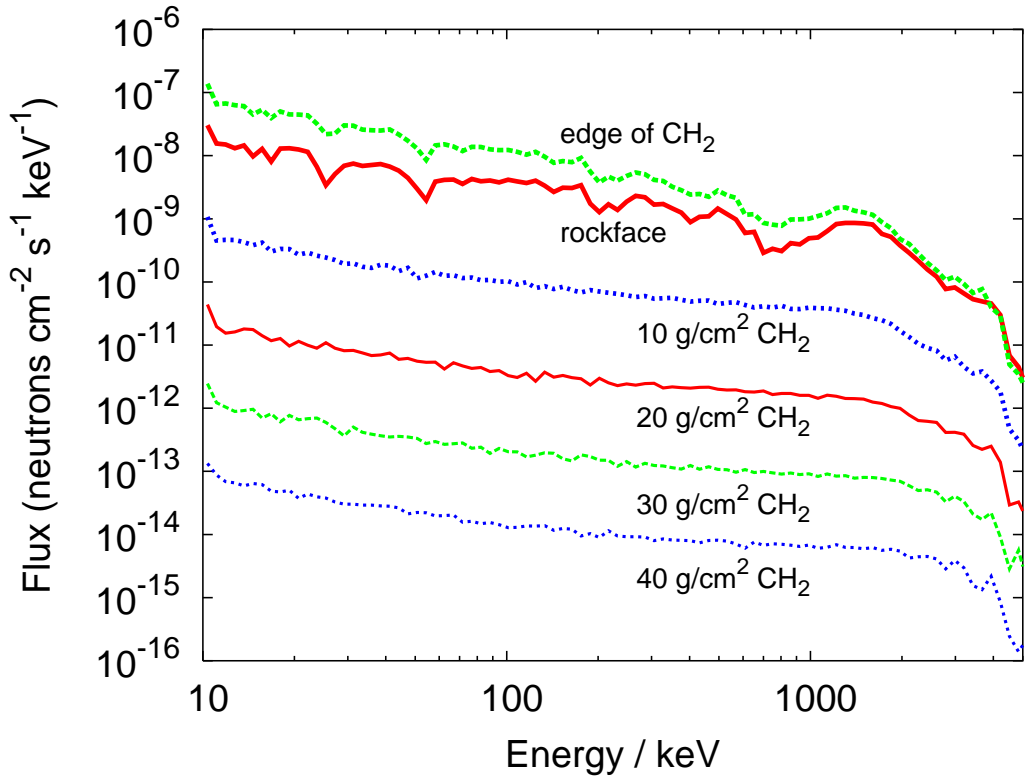,width=15cm}
\caption{Simulated energy spectra of neutrons originating in the
surrounding rock, due to uranium and thorium contamination, and
travelling through different layers of CH$_{2}$ shielding. Thick solid
curve - rock/lab boundary; thick dashed line - lab/shielding boundary
(note an increase in the rate due to back-scattering of neutrons from
the walls); thick dotted line - after 10 g cm$^{2}$ CH$_{2}$; thin
solid line - after 20 g cm$^{2}$ CH$_{2}$; thin dashed line - after 30
g cm$^{2}$ CH$_{2}$; thin dotted line - after 40 g cm$^{2}$
CH$_{2}$.}\label{ShLayerFig}
\end{center}
\end{figure}

Two amounts of CH$_{2}$ were investigated, $30$ g cm$^{-2}$ and $40$ g
cm$^{-2}$, to evaluate the effectiveness of the shielding. The
shielding tested with this simulation was CH$_{2}$ with a density of
$1$ g cm$^{-3}$.  The neutron flux through the layers of shielding in
the $40$ g cm$^{-2}$ simulation are shown in Figure
\ref{ShLayerFig}. These results are similar (after re-scaling) to
other simulations carried out using higher uranium and thorium
contamination levels \cite{nxenon}. The suppression of neutron flux 
by the CH$_{2}$ shielding agrees within a factor of 3 with the 
early simulations \cite{vk02} carried out with the MCNP code \cite{mcnp},
although different geometries and input neutron spectra used in the
present work and in early simulations make accurate comparison
difficult.

Four modules were also positioned side by side for some simulation
runs and varying amounts of CH$_{2}$ shielding were placed in between
the vessels to study the change in background rates and the
requirements for shielding between modules.

\begin{figure}
\begin{center}
\epsfig{figure=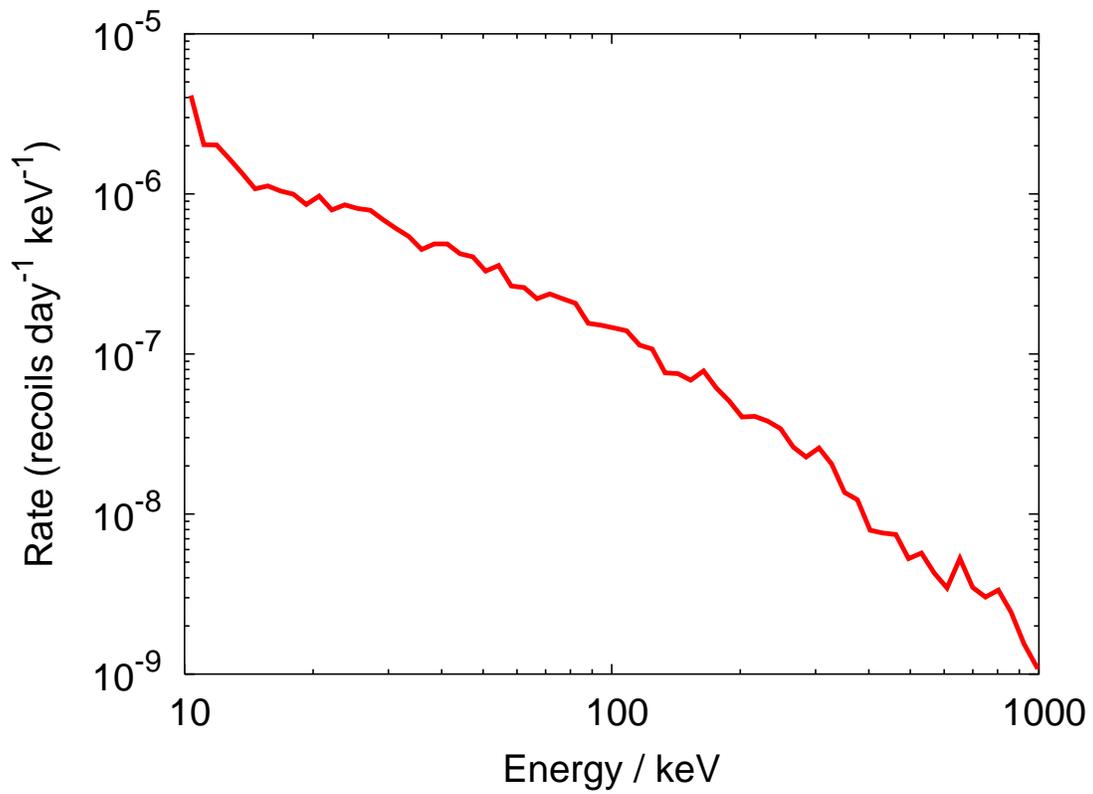,height=15cm,angle=-90}
\caption{Simulated energy spectrum of nuclear recoils in the 167 g of CS$_{2}$
  behind 40 g/cm$^2$ of CH$_2$ shielding produced by neutrons
  originating in the surrounding rock due to uranium and thorium
  contamination.}\label{rockRecoils}
\end{center}
\end{figure}

The results are summarised in Table \ref{rockRates} and the energy
spectrum of the nuclear recoils in a TPC behind $40$ g cm$^{-2}$ of
CH$_{2}$ is plotted in Figure \ref{rockRecoils}. Passive neutron
shielding of $30-40$ g cm$^{-2}$ of a hydrocarbon material is
required to suppress the nuclear recoil rate due to background
neutrons to well below $1$ per year for a gas TPC detector with
geometry as described above.  Neutron shielding between modules does
not seem to be necessary to sufficiently reduce the recoil rate due to
rock neutrons, although it does appear to have a small effect in
suppressing them when vessels are grouped together, since the total
amount of hydrocarbon material increases.

The results quoted here can be scaled to give expected recoil rates in
a similar detector with a different gas pressure or volume. A simple
linear scaling is reasonable up to about $4000$ torr, i.e. doubling the gas
pressure, or volume, doubles the recoil rate. Above circa $4000$ torr,
the probability of multiple scattering becomes non-negligible to
produce a bias in the results.

Some measurements suggest a harder neutron spectrum from rock than
obtained from SOURCES (see Ref. \cite{nxenon} for comparison). In this case 
5-10 g cm$^{-2}$ more of CH$_{2}$ may be required to ensure
similar suppression of neutrons.

\vspace{0.5cm}
{\large \bf 4. Muon-induced neutrons}
\vspace{0.3cm}

For the case of muon-induced neutrons 
the muon spectrum and angular distribution was simulated using the
MUSUN Monte Carlo code (see Ref. \cite{vakneutrons} for
description). Normalisation of the muon (and neutron) spectrum was
achieved using the measured value for the muon flux at the Boulby
Underground Laboratory: $(4.09 \pm 0.15) \times 10^{-8}$ cm$^{-2}$
s$^{-1}$, which corresponds to a rock overburden at vertical of
$2805 \pm 45$ m w.~e. \cite{muflux}. The quoted muon flux is 
for a spherical detector with unit cross-section. For
any other experimental site at similar depth the neutron flux can be
scaled roughly as the muon flux. Variation of muon energy spectrum
with depth can be accounted for by using the dependence of neutron
production on the mean muon energy as $\propto E^{0.79}$
\cite{vakneutrons}. Changes in the neutron flux due to different rock
composition around the laboratory (pure NaCl was used in this work)
can be estimated following the dependence of the neutron production on
the mean atomic weight of the rock as $\propto A^{0.76}$, although
large fluctuations from this relation are possible due to the
peculiarities in neutron production and scattering in different
materials \cite{vakneutrons}.

Muons were sampled on the surface of a cube of rock (NaCl) $20 \times
20 \times 20$ m$^{3}$. The laboratory cavern of size $6 \times 6
\times 5$ m$^{3}$ was placed inside the salt region at a depth of $10$
m from the top, and a distance of $7$ m from each vertical surface of
the cube. The cavern contained $20$, $30$ or 40 g cm$^{-2}$ thick
shielding made of hydrocarbon material (CH$_2$). A TPC design similar
to a single module described in the previous section was
used. Assuming a practical pressure of $160$ torr the fiducial mass
is equal to $167$ g. For most simulation runs the pressure was chosen
to be $4000$ torr ($25$ times the assumed operational TPC pressure)
and the results were re-scaled to the value of $160$ torr by dividing
the rate by $25$. Such an approach is valid since the probability of
double recoils at 4000 torr pressure is still $\ll 1$.  This was
checked by running simulations at lower and higher pressures.  At a
pressure $\le 4000$ torr more than 90\% of events are single nuclear
recoils and their numbers scale with density. At higher pressure the
fraction of double (and triple) nuclear recoil events increases with
density. The results obtained for $4000$ torr can be re-scaled to
lower values of pressure.

Simulations of muon propagation and interactions, development of
muon-induced cascades, neutron production, propagation and detection
were performed with FLUKA \cite{fluka}. Tests of muon-induced neutron
simulations with FLUKA can be found in \cite{vakneutrons,wang}. 
Recently the detailed comparison between FLUKA and GEANT4 for
muon-induced neutrons has been performed \cite{araujo}. It was shown
that the fluxes and spectra of fast neutrons agree within 20\%, and
the total neutron yields do not differ by more than a factor of 2
(depending on the material).

\begin{figure}[htb]
\begin{center}
\epsfig{figure=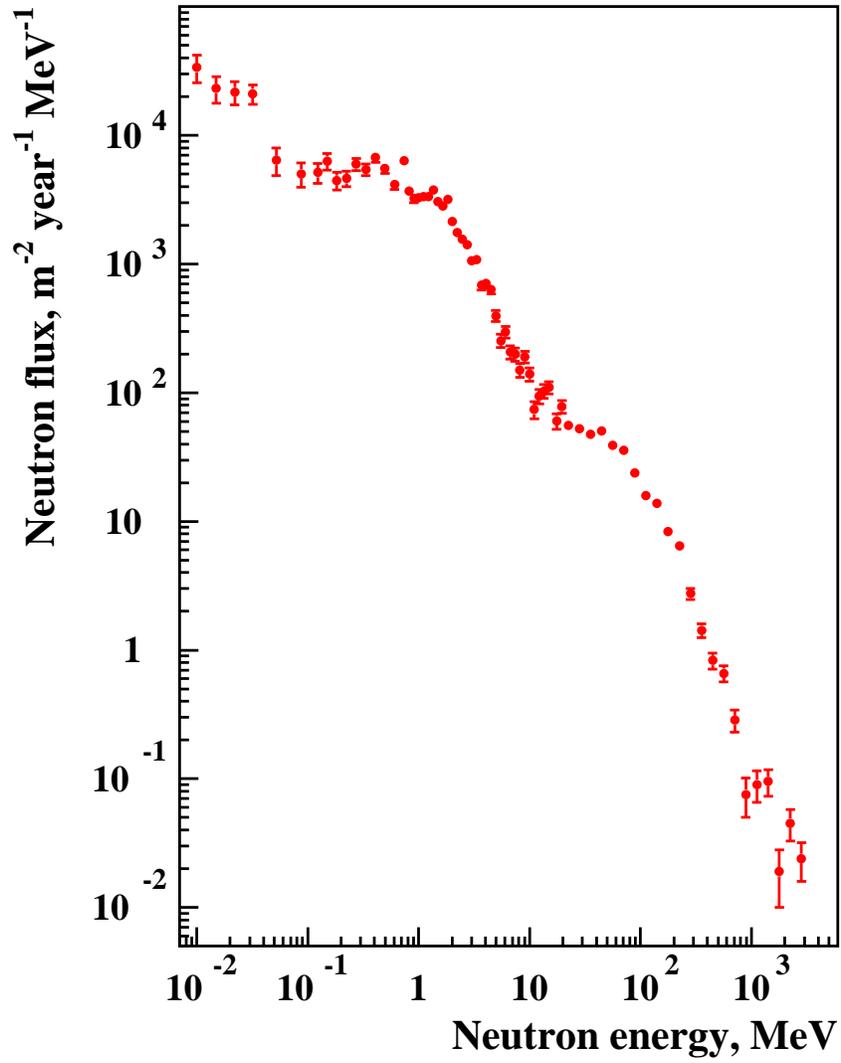,height=15cm}
\caption{Simulated energy spectrum of muon-induced neutrons
entering the TPC (vessel/gas boundary).}
\label{fig-nsptpc}
\end{center}
\end{figure}

The neutron production rate in NaCl at Boulby was found to be $7.6 \times
10^{-4}$ neutrons per muon per $1$ g cm$^{-2}$ of muon path. The total
neutron flux at the rock-cavern boundary is $3.4 \times 10^{-9}$
cm$^{-2}$ s$^{-1}$ above $0.1$ MeV if each neutron is counted as many
times as it enters the cavern (this takes into account neutrons
which cross the cavern, enter an opposite wall and are scattered back
into the cavern). Figure \ref{fig-nsptpc} shows the neutron spectrum
entering the TPC volume (vessel-gas boundary). 

Neutrons produced around the detector can cause nuclear recoils in the
gas that mimic WIMP-induced signals. Under the assumption that
DRIFT-type detectors can be made insensitive to gammas
\cite{smpdrift,dan,jakdrift}, all nuclear recoils (even those
accompanied by gammas, electrons or muons) are of potential danger to
the detector sensitivity.

\begin{figure}[htb]
\begin{center}
\epsfig{figure=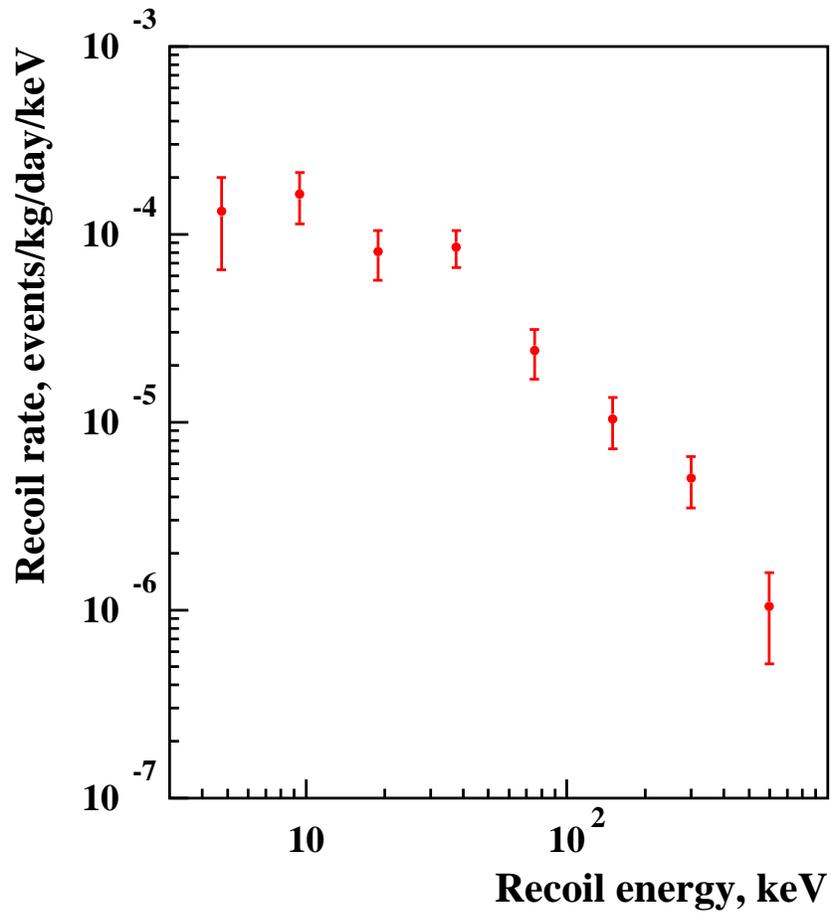,height=15cm}
\caption{Simulated nuclear recoil energy spectrum in a TPC filled with
  167 g of CS$_{2}$ from muon-induced neutrons. TPC is shielded with
  40 g/cm$^2$ of CH$_2$.} 
\label{fig-recsptpc}
\end{center}
\end{figure}

Figure \ref{fig-recsptpc} shows the energy spectrum of nuclear recoils
originated from muon-induced neutrons in a TPC filled with
CS$_{2}$. More than $90 \%$ are single recoils in $4000$ torr
gas. Obviously only single recoils will be detected at much smaller
pressures. Note, however, that FLUKA does not generate nuclear 
recoils below $19.6$ MeV neutron
energy realistically. Kerma factors, equivalent to the average energy
deposition, are calculated for neutron interactions. So, the energy
spectrum of recoils has larger systematic uncertainties than the total number
of neutron-induced events.

More than $120$ million muons were simulated in total for various gas
densities and shielding thicknesses, 
which corresponded to a live time of more than $500$ years for
a single module with $160$ torr. Only those runs with a density no
more than $4000$ torr were included in further analysis (about $80$
million muons -- $200$ years of running time with $160$ torr) to ensure
that multiple recoil events do not add more than $10 \%$ to the total
rate. The result of this simulation yields a 
total of $0.70 \pm 0.06$ nuclear recoils per year 
in a $167$ g fiducial mass module shielded with 40 g cm$^{-2}$ of CH$_2$. 
This also includes events where
nuclear recoils are in coincidence with any other form of energy
deposition in the detector not associated with nuclear recoils
(electrons, hadrons, muons etc.). The errors shown are purely
statistical. About a half of this rate  ($0.34 \pm 0.04$ per year)
corresponds to nuclear recoils only, without any other energy
deposition. Unfortunately, it is difficult to say without full Monte
Carlo of the detector response, whether a TPC will be in any way
sensitive to electrons, hadrons or muons, or what the energy threshold
could be, above which the energy deposition of these particles could be
detected. Moreover, since the simulations were performed at a pressure
that significantly exceeds the pressure for which results are quoted, 
the rate of stochastic interactions of photons, electrons, muons
and hadrons should be re-scaled to the required pressure, and the
resulting energy deposition from ionisation loss of electrons, muons
and hadrons should also be corrected accordingly. So, no definite
conclusion can be derived about the efficiency of rejecting neutron
events by looking at the associated energy deposition of accompanying
particles without full Monte Carlo of detector response at the actual 
operating pressure of a realistic detector. 
Keeping this in mind we present here two limiting cases, for
which either all neutron-induced nuclear recoils are the background,
or, assuming $100 \%$ efficiency of detecting other particles, only
nuclear recoils without any other charged particle in the fiducial
volume are of potential danger. About $0.16 \pm 0.03$ nuclear recoils
per year are expected to be within an energy range of interest for
dark matter searches ($10-50$ keV recoil energy). About half of
them ($0.08 \pm 0.02$ per year) are events with nuclear recoils only
(without any other particle entering the fiducial volume). This rate
also has an uncertainty associated with the treatment of nuclear
recoil energy in FLUKA.

Comparing the above rates of nuclear recoils with $0\%$ and $100\%$
efficiency for other particle detection, we conclude that for such a
TPC, the ability to detect ionisation from single-charge
particles could improve sensitivity. 
This can be achieved by lowering the threshold of the
detector and registering small signals on them if a TPC is triggered
by a nuclear-recoil-like pulse. Another way is to detect the sum of
energy depositions from all wires, in addition to the signal from each
wire. In this case a small signal on a single wire from ionisation
energy loss of a muon, for example, is below the threshold on a single
wire, but the total energy deposition on all wires can still be
visible.

Simulations were also carried out with $1$ cm thick vessel walls
(instead of a chosen standard thickness of $2$ cm, which gives a mass
of $1590$ kg of stainless steel). A decrease in neutron rate of circa
$10-20\%$ was found with $1$ cm walls, which does not significantly
exceed statistical fluctuations. This means that the vessel is not
expected to be the major contributor to the muon-induced neutron flux
at the detector.

To check the effect of the shielding the simulations were also carried
out with 20 and 30 g/cm$^2$ of CH$_2$. About a $30 \%$ increase in the
neutron-induced recoil rate was found with 20 g/cm$^2$ of CH$_2$,
whereas only small increase in recoil rate (not statistically
significant) was observed for 30 g/cm$^2$ of CH$_2$.

Most dark matter detectors also need lead (or other high-Z) shielding
to absorb gammas coming from the rock. A low pressure TPC, however, may
not need this since it can be made insensitive to gammas
\cite{smpdrift,dan}. Recently a new idea of using a TPC to detect
Kaluza-Klein axions was proposed \cite{ben}, which requires the
detection of low-energy gammas and, hence, suppression of gamma
background. To study the effect of lead shielding on neutron
production we added $10$ cm of lead between the rock and hydrocarbon
shielding (30 g/cm$^2$) and found an increase in the recoil rate of about 
$40-50\%$. This is due to the efficient production of neutrons in lead. Note
that by detecting accompanying cascade particles a TPC chamber could,
in principle, reject most of these events. 
By no means do we recommend installation of lead behind the
hydrocarbon shielding - this would result in a largely enhanced
production of neutrons in lead, which could then penetrate into the
main vessel left without hydrocarbon shielding. To reject those events
a powerful active veto with more than $99 \%$ efficiency for muon and
cascade detection is needed. More discussions about neutron production
in lead and hydrocarbon can be found in
Refs. \cite{nxenon,vakneutrons,wang}.

\vspace{0.5cm}
{\large \bf 5. Neutrons from detector components}
\vspace{0.3cm}

\begin{figure}
\begin{center}
\epsfig{figure=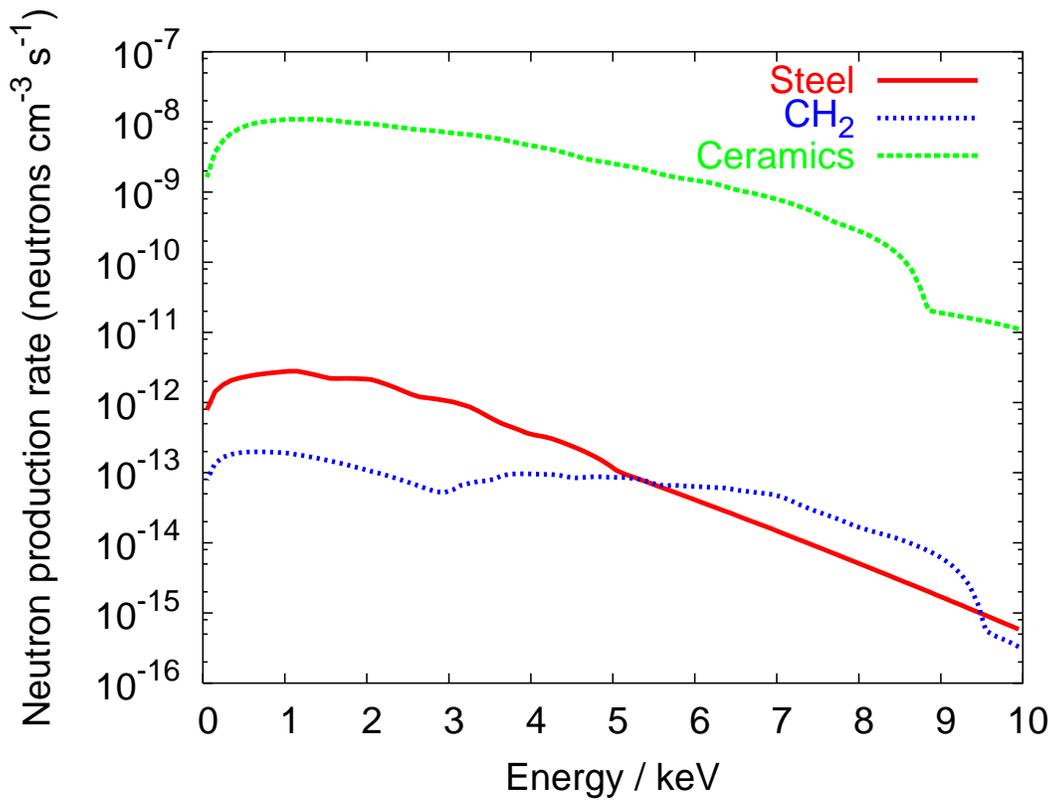,height=15cm,angle=-90}
\caption{Simulated energy spectra of neutrons produced in the detector
  components due to uranium and thorium contamination. Spectra are
  shown for the ceramics of the resistor chain, the stainless steel
  vessel and the CH$_{2}$ shielding surrounding the
  vessel.}\label{detNspecs}
\end{center}
\end{figure}

It is likely that the main limitation on the sensitivity of a dark 
matter detector is the rate of neutron background originating in the
detector itself. This is largely due to the difficulty of positioning
passive shielding within the detector and finding materials with low
radioisotope contaminations.  This makes simulation of these
backgrounds a very important task in determining the level of
radioisotope contamination that can reasonably be accepted in the
materials used to build the detector to ensure the sensitivity is not
significantly compromised. For the design of TPC used here, the
component materials expected to produce the largest neutron
backgrounds are the stainless steel vessel, the ceramic material of
the resistor chain and the hydrocarbon shielding. The neutron
production spectra generated by SOURCES for all three components are
shown in Figure \ref{detNspecs}.

In all runs simulating neutrons produced in detector components the
number of events processed was large enough to ensure that statistical
uncertainties were $\ll 1 \%$.

The levels of uranium and thorium contamination assumed to be present
in the stainless steel of the detector vacuum vessel were $0.5$ ppb
for each element. These values are reasonable for low
carbon stainless steel with low levels of radioisotopic impurities.
From the contamination values assumed one might expect a neutron
production rate of $7.01 \times 10^{-11}$ neutrons s$^{-1}$ cm$^{-3}$
in one $1590 \rm\, kg$ vessel ($2 \rm\, cm$ thick steel).

Simulations of neutrons produced from contaminants in the stainless
steel of the vessel were performed for a single module as well as for
four modules grouped together as described in section 2.  It was
assumed that shielding between the vessels would be of more importance
regarding the neutron background from the detector components because,
unlike that for the background from the surrounding rock, the fiducial
volume is not protected from the massive steel vessels by any passive
neutron shielding. It was found (see Table \ref{detectorRates}) that
the shielding between modules can be thin ($\sim 5 - 10$ g cm$^{-2}$) in
comparison to that required externally to shield from rock neutrons,
and that adding more shielding has very little effect on the
resulting recoil rates.

For the main element of stainless steel -- iron -- only the
($\alpha$,n) cross-section on $^{54}$Fe has been measured, hence we
used this cross-section for all iron isotopes. Although for some
isotopes in stainless steel the energy threshold for ($\alpha$,n)
reactions is quite low, the Coulomb barrier suppresses the
cross-section below about 7 MeV. Our calculations with SOURCES show
that spontaneous fission dominates the neutron production from uranium
in iron. We compared our results for a total neutron yield from U and
Th in iron and stainless steel with a calculation carried out by
Heaton \cite{heaton} for iron on the basis of the measurements by West
and Sherwood \cite{ws}. We found that for equal contamination levels
of U and Th, Heaton's yield for iron is $57 \%$ higher than our result
for iron. The spontaneous fission rate of U is the same in both cases,
but its contribution to the total yield is different for the two
calculations: spontaneous fission dominates in our case but is roughly
equal to the ($\alpha$,n) reaction contribution in Heaton's
estimate. According to the measurements \cite{ws} and our simulations
with SOURCES, stainless steel gives a slightly higher neutron yield
than pure iron, so for stainless steel our yield is about $70 \%$
smaller than was suggested by the measurements. We can interpret this
difference as an estimate for the systematic uncertainty in our
calculations. Unfortunately, the calculations of Heaton \cite{heaton}
cannot be used to generate recoil spectra since they do not contain
information about neutron energies and the cross-sections of
($\alpha$,n) reactions are not available from the measurements quoted
above.

The levels of uranium and thorium contamination assumed to be present
in the ceramic material of the resistor chain in the TPC were $500$
ppb and $2000$ ppb, respectively. These values were estimated from
measurements of several resistors \cite{ukdmc}.  The values are not
expected to be precise as they can vary significantly from one batch of
resistors to another, but they are reasonable estimated averages that
are, if anything, biased slightly high to provide upper limits on the
recoil rates produced.  The neutron production rate in ceramics with
these values of radioisotope content is $3.93 \times 10^{-7}$ neutrons
s$^{-1}$ cm$^{-3}$. The estimate of the number of resistors to be used
in a TPC detector of this design is $70$ and average resistor
dimensions are $3.7$ mm diameter and $10$ mm length, giving a total
mass of $22.6\rm\, g$ of ceramics.  Simulations were then run using
a single detector geometry surrounded by $40$ g cm$^{-2}$ of CH$_{2}$
shielding.

There were several measurements of the uranium and thorium contaminations
in the hydrocarbon materials used for neutron shielding \cite{ukdmc}. 
We used here the concentrations of $0.27$ ppb U and $0.05$ ppb Th, as an
approximate average of the measured values. The neutron production
rate calculated for this CH$_{2}$ material with the contamination
levels mentioned above is $4.5 \times 10^{-12}$ neutrons s$^{-1}$
cm$^{-3}$. A single module surrounded by $40 \rm\, g\, cm^{-2}$ of
CH$_{2}$ ($7.11 \rm\, m^{3}$) was simulated, with a total mass of
$7110\rm\, kg$.

\begin{figure}
\begin{center}
\epsfig{figure=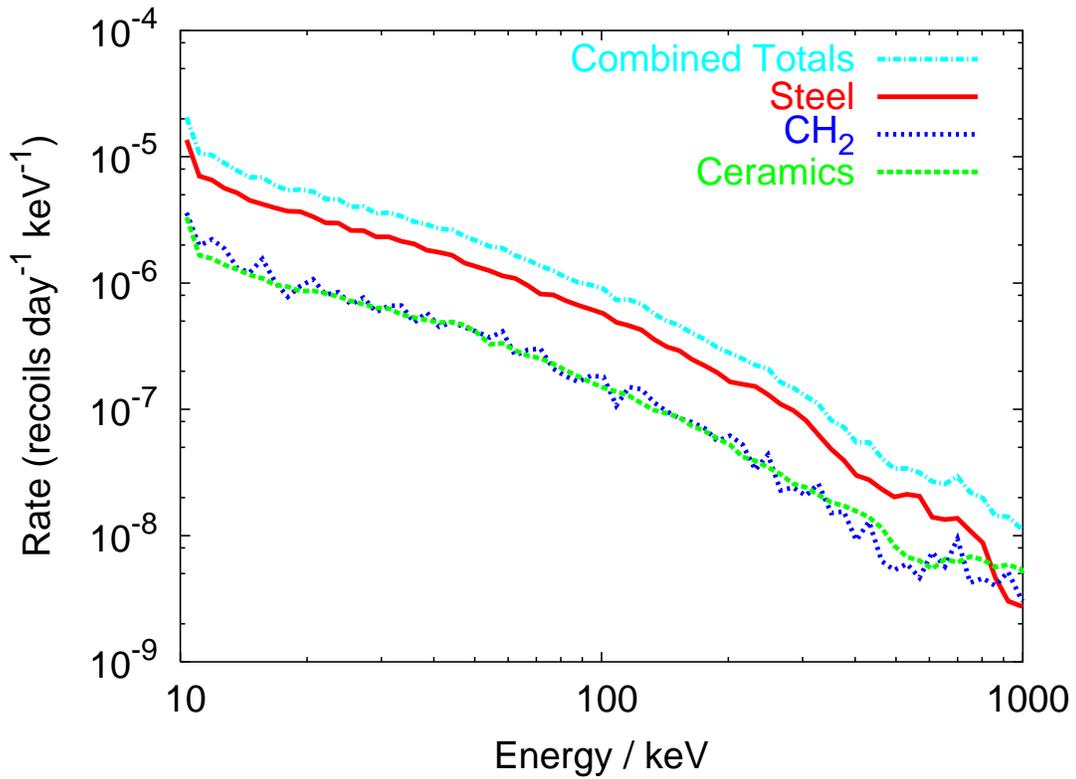,height=15cm,angle=-90}
\caption{Simulated energy spectra of nuclear recoils in the 167 g of CS$_{2}$
  produced by neutrons originating in detector components due to
  uranium and thorium contamination. Dotted curve - total combined
  recoil rate (steel + ceramics + CH$_{2}$); solid curve - recoil rate
  due to neutrons produced in the steel vessel; dashed curve - recoil
  rate due to neutrons produced in the CH$_{2}$ shielding; dash-dotted
  curve - recoil rate due to neutrons produced in the ceramics of the
  resistor chain.}\label{detRecoils}
\end{center}
\end{figure}

The nuclear recoil energy spectra produced by neutrons from these
three detector components are shown in Figure \ref{detRecoils}. The
resulting rates from the detector components are given in Table
\ref{detectorRates}. The contribution to the recoil rate at energies
between $10$ and $50$ keV due to neutron background from detector
components for a single detector as described here is $0.064$ recoils
per year. The recoil rate at energies between $10$ and $100$ keV due
to these backgrounds is 40\% higher. This rate can be suppressed by more
than an order of magnitude by installing 10 g/cm$^2$ of additional
CH$_2$ shielding just around the sensitive volume of the detector.

Another background source to be considered is radon. The associated
background is that of heavy nuclear recoils from radon decay products
(see also Ref. \cite{nxenon} for discussion), implanted on the surface
on the TPC vessel and other components.  These recoils, along with
alphas may contribute to the indistinguishable background events,
however, both radon associated and alpha events can be rejected if
they enter the fiducial volume from an edge, where active veto wires
are positioned on the MWPCs (most recoils and alphas will be produced
in the vessel walls or internal shielding).  Detailed simulation of
the detector response to alphas and nuclear recoils (including those
from detector internal parts) for an approved DRIFT II design is in
progress.

\vspace{0.5cm}
{\large \bf 6. Bounds on WIMP sensitivity of a low pressure TPC 
due to neutron background}
\vspace{0.3cm}

From the simulations of neutron background we can now investigate its
effect on the sensitivity of a low pressure gas TPC acting as a dark
matter detector. We assume that both gamma and alpha backgrounds are
fully suppressed and/or rejected. As mentioned above, gammas (via
electrons) induce typical signals that are below the energy threshold or
can be cut, giving a suppression of more than $10^{5}$
\cite{smpdrift,dan}. Alphas can be efficiently rejected as they are
produced mainly in the vessel walls and readout wires, have long
tracks and can be vetoed by wires close to the walls. Reduction of the
sensitive volume of the detector relative to the total volume of gas
is partly due to the necessity of rejecting alphas from the vessel
walls.

We consider a module described above with a fiducial mass of 167
g. The contributions to the nuclear recoil rate at 10-50 keV from
different sources are shown in Table \ref{table-n}. The thickness of
hydrocarbon shielding around the detector was assumed to be 40 g
cm$^{-2}$, reducing the recoil rate at 10-50 keV due to rock
neutrons to 0.01 events/year (column 2 in Table \ref{table-n}).

For muon-induced neutrons we consider here a case of $50\%$ efficiency
in rejecting nuclear recoils associated with any other charged
particle in the fiducial volume. This is a reasonable estimate for the
detector sensitivity. The nuclear recoil rate is shown in column 3.

For neutrons from detector components we present a case for a single
module surrounded by shielding (column 4). Contamination levels of 0.5
ppb were assumed for both U and Th in the stainless steel vessel,
which is a reasonable estimate made using a number of available
measurements.

It is obvious that the mass of a single module is not enough for the 
detector to be sensitive to the particle dark matter candidates 
favoured by SUSY models. A realistic detector will probably have 20 or 
60 modules with the fiducial masses of 3.33 and 10 kg, respectively.
The recoil rates for these detectors are shown in the 3rd and 4th 
rows of Table \ref{table-n}. We consider here modules separated by
the CH$_2$ shielding to reduce the neutron flux from detector
components.

It is clear that even for $10$ kg fiducial mass, the recoil rate at
$10-50$ keV from rock neutrons does not limit the detector
sensitivity if $40$ g cm$^{-2}$ of hydrocarbon shielding is installed
around the detector. Muon-induced neutrons are more significant giving
between $5$ and $10$ events per year (in case of $40$ g/cm$^{2}$ 
of CH$_2$) in a $10$ kg detector, depending
on the efficiency of single-charge particle detection ($7.2$ events in
Table \ref{table-n} for $50\%$ efficiency). This number can be reduced
by a factor of $10-20$ by installation of an active veto, for instance
a liquid scintillator, around the detector. Such a veto can also act
as a shielding against rock neutrons in the same way as any other
hydrocarbon material. Note that muon-induced neutrons effect the TPC
sensitivity to dark matter more than is the case for 
high density targets \cite{nxenon}. This is because the low pressure makes
the detector much less sensitive to gammas and electrons, coincidence
with which helps in rejecting nuclear recoils in 
liquid/solid materials. Another reason is the back-scattering of neutrons
from the internal walls of the vessel which
increases the total flux of neutrons through the target in TPC.

\begin{figure}[htb]
\begin{center}
\epsfig{figure=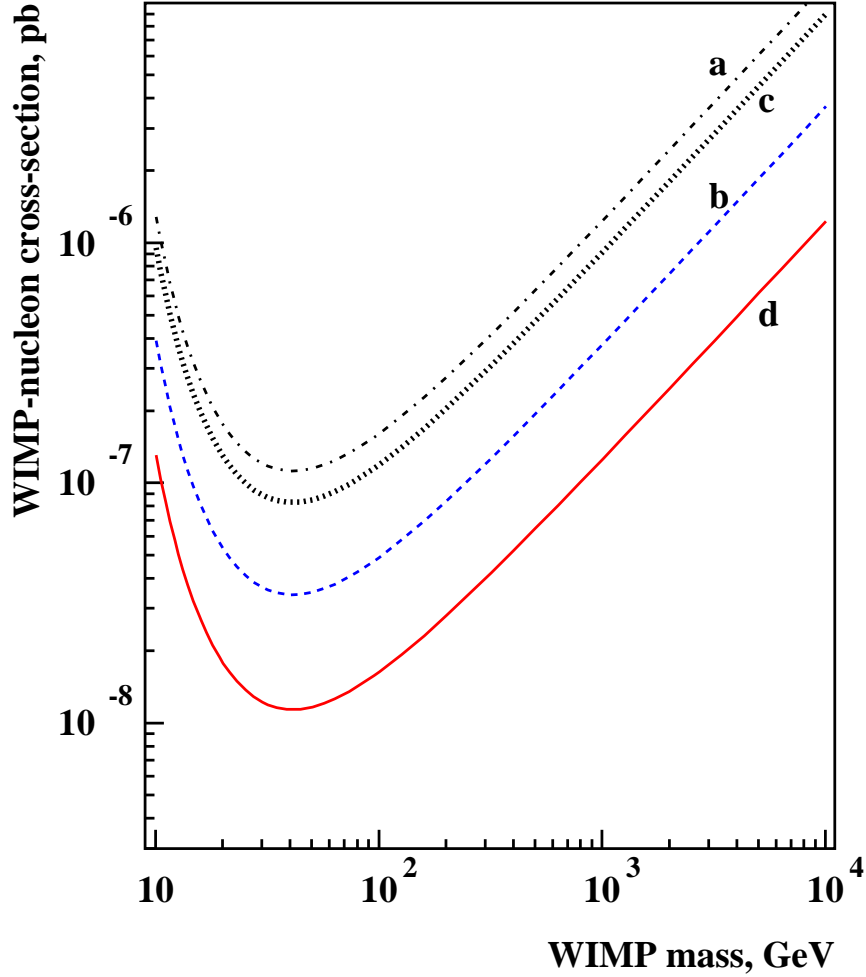,height=15cm}
\caption{Limitations on the sensitivity due to neutron backgrounds of a future
low-pressure gas TPC (one year running time and assuming 100\%
efficiency above 10 keV) for dark matter searches
assuming 100\% rejection of gamma and alpha induced events. For any
specific assumptions about the neutron-induced rate the parameter
space below the curve cannot be probed because of the neutron 
background (see text for details).
Dashed-dotted curve (a) -- 3.33 kg detector with 4 neutrons detected
in a year of running time; 
dashed curve (b) -- 3.33 kg detector with 0 events recorded; 
dotted curve (c) -- 10 kg detector with 12 events detected;
solid curve (d) -- 10 kg detector with 0 events recorded.
Nuclear recoils from alpha decays in the materials around the
fiducial volume were 
assumed to be vetoed by wires adjacent to the walls, whereas 
the rate of recoils and alphas from MWPC and cathode wires was 
neglected.}
\label{fig-senstpc}
\end{center}
\end{figure}

The bounds on the sensitivity at $90\%$ confidence level (CL) of a low
pressure TPC detector from  neutron background are shown in Figure
\ref{fig-senstpc} for $1$ year of running time. They were calculated
using the same procedure as for the case of large-scale xenon detector 
\cite{nxenon}. In all cases the
energy range of $10-50$ keV was used for analysis. In a real
experiment, the measured efficiency of the detector as a function of
energy should be used to set limits. Nuclear recoils from alpha decays 
in the materials around the fiducial volume were 
assumed to be vetoed by wires adjacent to the walls, whereas 
the rate of recoils and alphas from MWPC and cathode wires was neglected.

The upper solid curve shows the
best limit, which can be achieved with a $3.33$ kg detector ($20$
standard modules) in one year of running, assuming $4$ events detected
from muon-induced neutrons and detector components (upper limit of $8$
events at $90\%$ CL from \cite{pdg}). This number corresponds to about
$50\%$ efficiency of rejecting events in coincidence with any other
charged particle in the sensitive volume. The number of events can be
reduced to less than $1$ per year by installing a muon/neutron veto
around the detector. Assuming $0$ events detected we can plot a
sensitivity curve corresponding to the upper limit of $2.44$ at $90\%$
CL (dotted line). 

For a $10$ kg detector we can take $12$ events per year detected from 
muon-induced neutrons and detector components.
This gives an upper limit of $17.8$ events per year
at $90\%$ CL and the limit shown by dashed curve. Again, assuming an
efficient active veto and smaller U/Th levels in the vessel, this
number can be cut down to well below $1$ event/year and the limit will
be as shown by the lower solid curve (assuming $0$ events detected).

\vspace{0.5cm}
{\large \bf 7. Summary and conclusions}
\vspace{0.3cm}

We conclude that to suppress the nuclear recoil rate due to neutrons
produced in the rock to a reasonable level in a low pressure CS$_2$ TPC, 
$30-40$ g cm$^{-2}$
of hydrocarbon shielding is required around the modules.  It is also
advisable, due to the rate of neutrons produced in the steel vacuum
vessels, to place at least $5$ g cm$^{-2}$ of hydrocarbon shielding
between modules that are positioned side by side.  The total recoil
rate expected in a single module, in an array of $20$ modules and in
an array of $60$ modules due to the neutron backgrounds described
above are $0.19$, $3.9$ and $11.6$ events per year respectively,
assuming $50 \%$ detection efficiency for single-charge particles.
Adding low-background hydrocarbon shielding around the sensitive
volume of the detector to protect it from neutrons from detector 
components, and installing an active veto around the detector against
muons and their secondaries, would reduce the background rate by
an order of magnitude.
This would allow the sensitivity of $(1-3) \times 10^{-8}$ pb to
spin-independent WIMP-nucleon cross-section to be achieved with a $10$
kg TPC.

\vspace{0.5cm}
{\large \bf 8. Acknowledgements}
\vspace{0.3cm}

This work was performed in affiliation with the UK Dark Matter
Collaboration (University of Edinburgh, Imperial College London,
Rutherford Appleton Laboratory and University of Sheffield) and
contributes to collaboration-wide simulation efforts. We also
acknowledge the work performed by our colleagues at Occidental College,
L.A., and Temple University, Philadelphia, especially
Prof. D.P. Snowden-Ifft and Prof. C.J. Martoff.  This work is funded
by PPARC. We also acknowledge the funding from EU FP6 programme --
ILIAS.

\pagebreak
\begin{table}[h!]
 \caption{Recoil rates expected in a single gas TPC module due to
 neutrons from the surrounding rock from simulations of single and
 multiple modules. Event rates with and without hydrocarbon neutron
 shielding are shown.}\label{rockRates}
\begin{center}
\begin{tabular}{|p{6.5cm}|p{3.5cm}|p{3.5cm}|}
\hline
Description of simulation & Nuclear recoil rates in a single TPC
module per year at $10-50$ keV & Ratio of 10-50 keV rate to 10-100
keV rate\\
\hline
Unshielded single module & \hspace{0.65cm} $2.67 \times 10^{3}$ &
\hspace{1.2cm} $0.81$ \\
\hline
Single module, $30$ g cm$^{-2}$ external CH$_{2}$ shielding &
\hspace{0.8cm}$1.70 \times 10^{-1} $ & \hspace{1.2cm} $0.73$ \\
\hline
Single module, $40$ g cm$^{-2}$ external CH$_{2}$ shielding &
\hspace{0.8cm}$1.10 \times 10^{-2}$ & \hspace{1.2cm} $0.70$ \\
\hline
4 modules, no intermediary shielding, $40$ g cm$^{-2}$ external CH$_{2}$ & 
\hspace{0.8cm}$1.00 \times 10^{-2}$ & \hspace{1.2cm} $0.73$ \\
\hline
4 modules, $40$ g cm$^{-2}$ CH$_{2}$ between modules, $10$ g cm$^{-2}$ external
CH$_{2}$ & \hspace{0.8cm}$7.97 \times 10^{-3}$ & \hspace{1.2cm} $0.71$ \\
\hline
\end{tabular}
\end{center}
\end{table}

\pagebreak
\begin{table}[h!]
 \caption{Recoil rates expected in a single gas TPC module due to
 neutrons from detector components from simulations of single and
 multiple modules. All simulations were performed with $40$ g
 cm$^{-2}$ external CH$_{2}$ shielding.}\label{detectorRates}
\begin{center}
\begin{tabular}{|l|p{6.0cm}|p{4.0cm}|}
\hline
Detector component & Description of simulation & Nuclear recoil rates
in a single TPC module per year at 10-50 keV \\
\hline
Steel vessel & Single module & \hspace{1.0cm}$4.14 \times 10^{-2}$ \\
\cline{2-3}
& 4 modules, no intermediary shielding & \hspace{1.0cm}$6.67 \times
10^{-2}$ \\
\cline{2-3}
& 4 modules, $5$ g cm$^{-2}$ intermediary CH$_{2}$ shielding &
\hspace{1.0cm}$4.57 \times 10^{-2}$ \\
\cline{2-3}
& 4 modules, $10$ g cm$^{-2}$ intermediary CH$_{2}$ shielding &
\hspace{1.0cm}$4.23 \times 10^{-2}$ \\
\cline{2-3}
& 4 modules, $20$ g cm$^{-2}$ intermediary CH$_{2}$ shielding &
\hspace{1.0cm}$4.20 \times 10^{-2}$ \\
\hline
Resistor ceramics & Single module & \hspace{1.0cm}$1.09 \times
10^{-2}$ \\
\hline
CH$_{2}$ shielding & Single module & \hspace{1.0cm}$1.13 \times
10^{-2}$ \\
\hline
\end{tabular}
\end{center}
\end{table}

\pagebreak
\begin{table}[h!]
\caption{ Neutron background rates per year at 10-50 keV 
recoil energies from different sources
in a 167 g, 3.33 kg and 10 kg TPC (40 g/cm$^2$ of CH$_2$ shielding
against rock neutrons was assumed; see text for details).}
\begin{center}
\begin{tabular}{|c|c|c|c|c|}
\hline
Detector mass & 
\multicolumn{4}{c|}{Nuclear recoil rates per year at 10-50 keV}\\
 kg & Rock & Muons & Detector & Total \\
\hline
 0.167 & 0.01 & 0.12 & 0.06 & 0.19 \\
\hline
 3.33 & 0.2 & 2.4 & 1.3 & 3.9 \\
\hline
 10.0 & 0.6 & 7.2 & 3.8 & 11.6 \\
\hline
\end{tabular}
\end{center}
\label{table-n}
\end{table}

\pagebreak

\pagebreak

\pagebreak

\pagebreak

\pagebreak

\pagebreak

\pagebreak

\pagebreak

\pagebreak

\pagebreak

\pagebreak

\end{document}